\documentclass[aps,pra,twocolumn,superscriptaddress,showpacs]{revtex4}

\usepackage{graphicx}

\begin{document}

\title{Bose-Einstein Condensation of an Ytterbium Isotope}

\author{Takeshi Fukuhara}
\affiliation{Department of Physics, Graduate School of Science, Kyoto University, Kyoto 606-8502, Japan}

\author{Seiji Sugawa}
\affiliation{Department of Physics, Graduate School of Science, Kyoto University, Kyoto 606-8502, Japan}

\author{Yoshiro Takahashi}
\affiliation{Department of Physics, Graduate School of Science, Kyoto University, Kyoto 606-8502, Japan}
\affiliation{CREST, Japan Science and Technology Agency, Kawaguchi, Saitama 332-0012, Japan}

\date{\today}

\begin{abstract}
We report the observation of a Bose Einstein condensate in a bosonic isotope of ytterbium ($^{170}$Yb). More than $10^6$ atoms are trapped in a crossed optical dipole trap and cooled by evaporation. Condensates of approximately $10^4$ atoms have been obtained.  From an expansion of the condensate, we have extracted the scattering length $a_{170}=3.6 \pm 0.9$ nm. 
\end{abstract}


\maketitle
Bose-Einstein condensation (BEC) of dilute atomic gases has been extensively investigated, which results in a deeper understanding of quantum gases. So far, BEC has been observed in ten species: H \cite{Fried98}, $^4$He \cite{Robert01}, $^7$Li \cite{Bradley95}, $^{23}$Na \cite{Davis95}, $^{41}$K \cite{Modugno01}, $^{52}$Cr \cite{Griesmaier05}, $^{85}$Rb \cite{Cornish00}, $^{87}$Rb \cite{Anderson95}, $^{131}$Cs \cite{Weber03}, and $^{174}$Yb \cite{Takasu03b}. One of the most exciting current directions in this field is to investigate Bose-Fermi and Bose-Bose degenerate mixtures. For example, the stability \cite{Modugno02a, Ospelkaus06}, the phase separation \cite{Hall98}, collective dynamics \cite{Modugno02b}, and strongly correlated systems in an optical lattice \cite{Gunter06} are studied. It is therefore still important to create a BEC in other atomic species or other isotopes because it might enable us to generate new interesting degenerate mixtures for various studies.

Ytterbium (Yb) is expected to play an important role in these studies because it has seven stable isotopes: five bosonic isotopes with nuclear spin $I=0$ ($^{168,170,172,174,176}$Yb) and two fermionic isotopes, $^{171}$Yb with $I=1/2$ and $^{173}$Yb with $I=5/2$. Following the realization of a BEC in $^{174}$Yb atoms, a degenerate Fermi gas of $^{173}$Yb atoms was achieved \cite{Fukuhara07}. A BEC in other Yb isotopes will enable us to perform various experiments on Bose-Fermi and Bose-Bose Yb degenerate mixtures. Especially, it is important to produce Bose-Bose mixtures because there are only a few investigations on two-species condensations \cite{Modugno02b}. Of particular interest is a study of two-species BECs in an optical lattice, which will exhibit a rich phase diagram \cite{Kuklov03, Paredes03, Zheng05, Alon06}. Moreover, the stability and phase diagrams of more than two BECs have been theoretically investigated \cite{Roberts06}. Yb atoms are suitable for these studies.

Generation of a BEC in an even isotope of Yb is also helpful in a recently proposed and demonstrated optical lattice clock using even isotopes of alkaline-earth-like atoms \cite{Takamoto05, Taichenachev06}. To eliminate the collisional frequency shifts and to obtain the largest signal-to-noise ratio, it is favorable to prepare each atom in each site of a three-dimensional optical lattice. This preparation can be realized by using the Mott insulator transition after loading a BEC into optical lattices \cite{Greiner02}. While the BEC of $^{174}$Yb was already obtained, the production of a BEC in another bosonic isotope is useful because the simultaneous operation of two optical lattice clocks enables us to evaluate the stability as well as the accuracy of optical lattice clocks at the uncertainty of the $10^{-16}$ level and beyond, where no working standard exists. 

In this article, we report all-optical generation of a BEC in $^{170}$Yb atoms. After loading the atoms from a magneto-optical trap (MOT) into a crossed-beam optical dipole trap, evaporative cooling is performed by lowering the depth of the trap. In 12 s, the trap depth along the vertical direction is reduced below 2 $\mu$K, where the bimodal distribution of the velocity and an anisotropic expansion after release from the trap are observed as evidences of the BEC transition. From the expansion of the BEC, we deduce the scattering length of $^{170}$Yb. 

\begin{figure}
\begin{center}
\includegraphics[width=\linewidth]{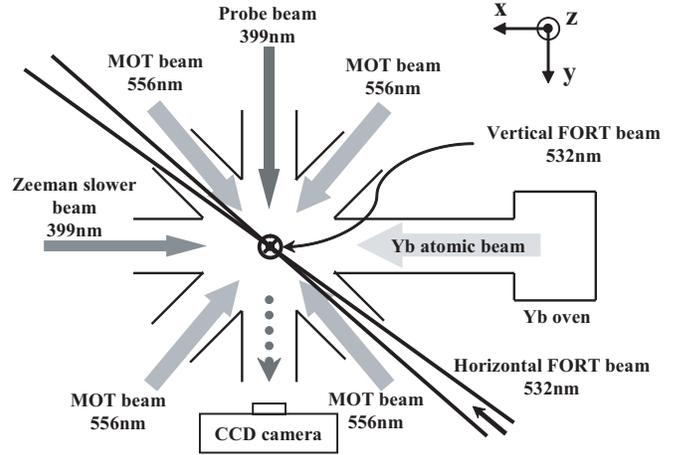}
\end{center}
\caption{Top view of the experimental setup for laser cooling, optical trapping, and probing of $^{170}$Yb atoms. Two MOT beams along the vertical z direction are not shown for clarity. We note that the probe beam used for detection is not perpendicular to the horizontal FORT beam. \label{setup}}
\end{figure}

The experimental apparatus used for this work is essentially the same as our previous experiments \cite{Takasu03b, Fukuhara07} and depicted in Fig. \ref{setup}. 
Yb atoms in a thermal beam generated from an oven at 375 $^{\circ }$C are decelerated by a Zeeman slower with a strong transition ($^1$S$_0-^1$P$_1$; the wavelength of 399 nm and the linewidth of 29 MHz) and then loaded into the MOT with an intercombination transition ($^1$S$_0-^3$P$_1$; the wavelength of 556 nm and the linewidth of 182 kHz) \cite{Kuwamoto99}. Since the natural abundance of $^{170}$Yb is relatively small (3.05\%), it takes typically 120 s to collect about $10^7$ atoms in the MOT, which is sufficient for the subsequent evaporation to BEC, while typically 10 s is needed in the case of $^{174}$Yb, whose natural abundance is 31.8\%.  

The laser-cooled $^{170}$Yb atoms are transferred to the crossed far-off-resonance trap (FORT) \cite{Takasu03a} with horizontal and vertical beams which are left on throughout the laser cooling process. The beams are independently produced by two 10 W diode-pumped solid-state lasers at 532 nm. The $1/e^2$ beam radii at the crossing point are 14 $\mu$m (horizontal beam) and 86 $\mu$m (vertical beam). At this loading phase, each beam has a power of 5.2 W, and the beams generate the potential depths of approximately 650 $\mu$K (horizontal beam) and 18 $\mu$K (vertical beam). Since the trap depth of the horizontal beam is much deeper than that of the vertical one, $1.6\times 10^6$ atoms at 75 $\mu$K are trapped mainly in the horizontal FORT. Atom numbers and temperatures of the gases are measured using an absorption imaging technique. The trapping beams are turned off within a few hundred ns. After a certain expansion time, the released gas is illuminated by 100 $\mu$s probe beam pulse resonant with the strong $^1$S$_0-^1$P$_1$ transition. 

The collisional properties are crucially important as to whether the atoms can be cooled to quantum degeneracy by evaporative cooling. In the case of Yb atoms, there is no binary inelastic collisions resulting from electron spins, and thus the binary collisions are only due to elastic collisions where the scattering length is the most important parameter. While the scattering length of another isotope of $^{174}$Yb has been precisely determined to be 5.5 nm from a photoassociation experiment \cite{Enomoto07}, there was no experimental and theoretical information on the scattering length of $^{170}$Yb \cite{Kitagawa07}. For this reason, we optimized the sequence of evaporation by variously changing the time constant. For a single-beam optical trap, the relation between the elastic collision rate and the time constant of forced evaporation has been discussed \cite{O'Hara01}. However, in our case of the crossed-beam optical trap, the process of evaporation is complicated and thus quantitative discussion is difficult.

Evaporative cooling is carried out by ramping down the intensity of the horizontal beam, while keeping the vertical FORT power constant. When the potential depth of the horizontal FORT is reduced to approximately the same as that of the vertical FORT, almost all the atoms are trapped in the crossed region. Further decrease of the horizontal FORT beam intensity leads to evaporation in the crossed region. Finally, the potential depth along the vertical direction, which is considerably affected by the gravity at this stage of evaporation, is reduced below 2 $\mu$K ($\sim$50 mW of power in the horizontal beam), where degenerate gases appear. The total evaporation time optimized as described above is 12 s, which is twice longer than that for $^{174}$Yb. From this fact, it is deduced that the absolute value of the scattering length of $^{170}$Yb is smaller than 5.5 nm for $^{174}$Yb.

\begin{figure}
\begin{center}
\includegraphics[width=\linewidth]{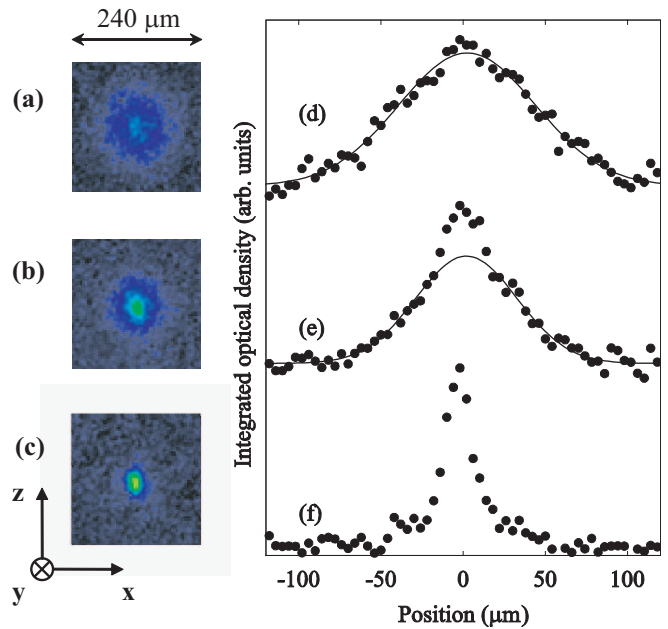}
\end{center}
\caption{(color online). Absorption images (a)$-$(c) and density distributions integrated over the vertical z direction (d)$-$(f) of $^{170}$Yb atoms after 10 ms of the free expansion: (a), (d) thermal cloud, $T=360$ nK; (b), (d) bimodal distribution, $T=200$ nK; (c), (f) almost pure condensate with $7 \times 10^3$ atoms. The solid lines show the Gaussian fits to the thermal components. \label{170bec}}
\end{figure}

The BEC phase transition has been observed in the density distributions after 10 ms of a free expansion. At the power of 70 mW in the horizontal FORT beam, the atom cloud has an isotropic Gaussian distribution expected for a thermal cloud [Fig. \ref{170bec}(a), (d)]. The corresponding temperature is $T=360$ nK. When the power of the horizontal beam is lowered down to approximately 50 mW, a bimodal momentum distribution appears [Fig. \ref{170bec}(b), (e)]. The temperature $T=200$ nK is determined from a Gaussian fit to the wings of the thermal cloud. As the power is decreased further, the central anisotropic component of a BEC increases. At the final power of 40 mW, the cloud is an almost pure condensate with approximately $10^4$ atoms [Fig. \ref{170bec}(c), (f)]. Even after reducing the power of the horizontal beam to perform evaporation, the radial confinement of the horizontal FORT is stronger than that of the vertical FORT, and thus the condensate released from the optical trap rapidly expands in the radial direction of the horizontal FORT beam due to the release of mean field energy. Since the probe beam propagates horizontally at an angle of approximately 50$^{\circ }$ with respect to the horizontal FORT beam (Fig. \ref{setup}), the observed absorption image of the condensate for expansion times longer than a few millisecond is elongated in the vertical z direction [Fig. \ref{170bec}(c)].

\begin{figure}
\begin{center}
\includegraphics[width=\linewidth]{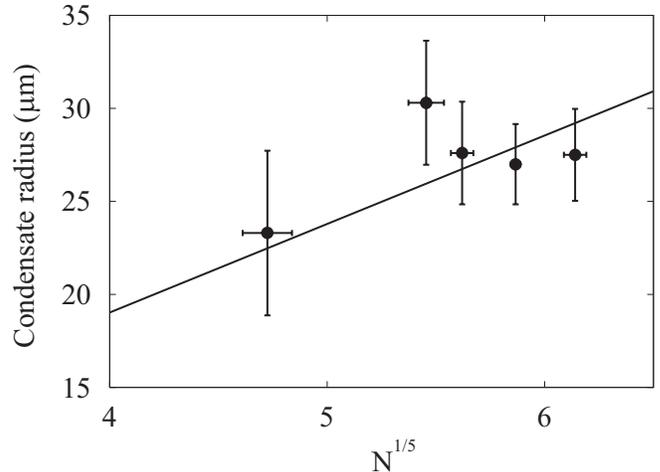}
\end{center}
\caption{Measured condensate radius along the vertical z direction after the free expansion of 10 ms, as a function of the number of condensate atoms  $N^{1/5}$. The solid line is a linear fit through the origin. The slope of the line extracts the scattering length $a_{170} = 3.6 \pm 0.9$ nm. \label{expansion}}
\end{figure}

The scattering length is an important parameter in the study of ultracold dilute atoms. It dominates the dynamics and stability of quantum degenerate gases. For this reason, it is significant to obtain information on the scattering length of $^{170}$Yb. We evaluate the scattering length of the $^{170}$Yb from the size of the observed condensate. In general the free expansion of the condensate is described by the Gross-Pitaevskii equation, and within the Thomas-Fermi approximation it is characterized by the set of equations for scaling factors, which can be solved easily \cite{Castin96}. In this approximation the condensate density is a time-dependent inverted parabola
\begin{equation}
n(\vec{r},t)=n_0(t) \left[ 1-\Sigma_{i=1}^{3} r_i/\{ \lambda_i(t) R_i \} \right], \label{TF}
\end{equation}
when the right hand side is positive and $n(\vec{r},t)=0$ otherwise. Here $i=$ 1, 2, and 3 corresponds to the spatial coordinates x, y, and z, respectively. The $\lambda_i(t)$ are scaling factors for the condensate radii, the $R_i$ are the Thomas-Fermi radii of the condensate in the trap, and the $n_0(t)$ is the peak density. The scaling factors $\lambda_i(t)$ satisfy
\begin{equation}
\ddot{\lambda_i}=\frac{\omega_i^2}{\lambda_i \lambda_1 \lambda_2 \lambda_3} \qquad ( i = 1, 2, 3 ),
\end{equation}
where the $\omega_i$ are the final trapping frequencies before release. The initial conditions are $\lambda_i(0)=1$ and $\dot{\lambda_i}(0)=0$. We note that the scaling factors can be calculated only from the trap frequencies and the expansion time by numerically solving these equations. The Thomas-Fermi radii are 
\begin{equation}
R_i = \frac{1}{\omega_i} (15 \frac{\hbar ^2}{m^2} \omega_1 \omega_2 \omega_3 N a)^{1/5}, 
\end{equation}
where $m$ is the atomic mass, $N$ is the number of condensate atoms, and $a$ is the scattering length. Therefore, the condensate radii $R_i(t) = \lambda_i(t) R_i$ after an arbitrary expansion time $t$ are proportional to $(N a)^{1/5}$, and the constants of proportionality depend only on the trap frequencies,  the expansion time, and the atomic mass. We have examined the condensate radii after the expansion for different atom numbers, and evaluated the scattering length of $^{170}$Yb. We plot the condensate radius along the vertical z direction after 10 ms of the expansion, as a function of the number of condensate atoms $N^{1/5}$ and fit the data with a straight line through the origin (Fig. \ref{expansion}). 
From the slope of the line we have extracted the scattering length $a_{170} = 3.6 \pm 0.9$ nm. This relatively small value of the scattering length of $^{170}$Yb is consistent with the relatively long evaporation time required to achieve the BEC of $^{170}$Yb compared with the case of $^{174}$Yb, as mentioned above. It is also impressive that this value is in good agreement with the quite recent result obtained by a completely different method \cite{Kitagawa07}.

This successful production of a $^{170}$Yb BEC is an important step to investigate a $^{170}$Yb-$^{174}$Yb BEC-BEC mixture and a $^{170}$Yb-$^{173}$Yb Bose-Fermi degenerate mixture. It is worth noting that the inter-species scattering length between $^{170}$Yb and $^{174}$Yb was found to have a large and negative value by using mass scaling \cite{Kitagawa07}, and thus we can observe collapse phenomena in this BEC-BEC mixture. The experimental procedure to generate a $^{170}$Yb BEC is also applicable to realization of a BEC in $^{168}$Yb, whose natural abundance is extremely small (0.13\%) and whose scattering length is positive \cite{Kitagawa07}. Moreover, the positive inter-species scattering length between $^{168}$Yb and $^{170}$Yb enables us to investigate a stable BEC-BEC mixture.

In conclusion, we have succeeded in the formation of a BEC in $^{170}$Yb atoms. We have determined the scattering length $a_{170} = 3.6 \pm 0.9$ nm from the expansion of the condensate. We emphasize that the straightforward application of the $^{170}$Yb BEC opens up studies of a $^{170}$Yb-$^{174}$Yb BEC-BEC mixture and also a $^{170}$Yb-$^{173}$Yb Bose-Fermi degenerate mixture.

We acknowledge Y. Takasu and S. Uetake for their experimental assistances and helpful comments. This research was partially supported by Grant-in-Aid for Scientific Research of JSPS (18043013, 18204035), SCOPE-S, and 21st Century COE "Center for Diversity and Universality in Physics" from MEXT of Japan.

\end{document}